\documentclass[twocolumn, preprintnumbers, showpacs, nofootinbib, aps, prl]{revtex4-1}
\usepackage{graphicx}

\newcommand{\bi}{\bibitem}
\newcommand{\be}{\begin{eqnarray}}
\newcommand{\ee}{\end{eqnarray}}

\begin{document}

\title{Constraint on the quadrupole moment of super-massive black hole candidates from the estimate of the mean radiative efficiency of AGN}

\author{Cosimo Bambi}
\email{cosimo.bambi@ipmu.jp}

\affiliation{
Institute for the Physics and Mathematics of the Universe, 
The University of Tokyo, Kashiwa, Chiba 277-8583, Japan}

\date{\today}

\preprint{IPMU11-0013}

\begin{abstract}
The super-massive objects at the center of many galaxies are
commonly thought to be black holes. In 4-dimensional general 
relativity, a black hole is completely specified by its mass 
$M$ and by its spin angular momentum $J$. All the higher
multipole moments of the gravitational field depend in a 
very specific way on these two parameters. For instance, the 
mass quadrupole moment is $Q = - J^2/M$. If we can estimate 
$M$, $J$, and $Q$ for the super-massive objects in galactic 
nuclei, we over-constrain the theory and we can test the black 
hole hypothesis. While there are many works studying how this 
can be done with future observations, in this paper 
a constraint on the quadrupole moment of these objects is obtained by 
using the current estimate of the mean radiative efficiency of 
AGN. In terms of the anomalous quadrupole moment $q$, the bound 
is $-2.01 < q < 0.14$.
\end{abstract}


\maketitle


{\it Introduction --}
Today we believe that the final product of the gravitational
collapse is a black hole (BH) and we have robust observational 
evidences of the existence of $5 - 20$~Solar mass compact 
objects in X-ray binary systems~\cite{bh1} and of $10^5 - 
10^9$~Solar mass objects at the center of many galaxies~\cite{bh2}.
All these objects are interpreted as BHs because they cannot 
be explained otherwise without introducing new physics. The 
stellar-mass objects in X-ray binary systems are too heavy to 
be neutron or quark stars for any reasonable equation of 
state~\cite{rr}. At least some of the super-massive objects 
in galactic nuclei are too heavy, compact, and old to be 
clusters of non-luminous bodies~\cite{maoz}. However, there 
are no direct observational evidences that they have an event 
horizon~\cite{abra}, while there are theoretical arguments 
suggesting significant deviations from the classical picture~\cite{new}.

In 4-dimensional general relativity, (uncharged) BHs are described 
by the Kerr solution and are completely specified by two parameters: 
the mass, $M$, and the spin angular momentum, $J$. The condition
for the existence of the event horizon is $|a_*| \le 1$, where
$a_*=J/M^2$ is the dimensionless spin parameter. The fact that a
BH has only two degrees of freedom is known as ``no-hair'' 
theorem~\cite{thm} and implies that all the mass moments, 
$M_{\it{l}}$, and all the current moments, $S_{\it{l}}$, of the
gravitational field can be written in terms of $M$ and $J$ by the 
following simple formula:
\be
M_{\it{l}} + i S_{\it{l}} = M
\left(i\frac{J}{M}\right)^{\it{l}} \, .
\ee
The first three non-trivial terms are the mass $M_0=M$, the
spin angular momentum $S_1=J$, and the mass quadrupole moment
$M_2 \equiv Q = -J^2/M$. On the contrary, for a generic 
compact object $M_{\it{l}}$ and $S_{\it{l}}$
can assume any arbitrary value, but in case of reflection symmetry, 
all the odd $M_l$-moments and all the even $S_l$-moments are identically 
zero. As it was put forward by Ryan in~\cite{ry}, by measuring the
mass, the spin, and at least one more non-trivial moment of 
the gravitational field of a BH candidate, one over-constrains 
the theory and can test the Kerr BH hypothesis.

There is a whole line of research devoted to study how future
experiments will be able to measure the mass quadrupole moment
of BH candidates and thus test the nature of these objects. The 
most studied and promising approach is through the detection of
gravitational waves of the inspiral of a stellar-mass compact 
object into a super-massive BH candidate~\cite{gw}. Other 
proposals involve the observation of the BH shadow~\cite{shadow}, 
the possible discovery of a stellar-mass BH candidate with a 
radio pulsar as companion~\cite{radio}, and accurate measurements 
of stellar orbits at mpc distances from Sgr~A$^*$~\cite{will}.
There are also two proposals to constrain $Q$ with current
available X-ray data, by studying the K$\alpha$ iron line~\cite{iron}
and the disk's thermal spectrum~\cite{noi}. Only Ref.~\cite{noi} 
constrains $Q$ by considering the stellar-mass BH candidate
M33~X-7, but the analysis is based on a simplified model and
the bound is only meant as a qualitative guide for future more 
rigorous studies.

{\it Radiative efficiency of AGN --}
The energy radiated by a compact object as a consequence of the
accretion process is simply $L_{acc} = \eta \dot{M} c^2$, where
$\eta$ is the efficiency parameter, $\dot{M}$ is the mass
accretion rate, and $c$ is the speed of light. If the accreting
gas cannot radiate efficiently its gravitational energy and 
the compact object is capable of absorbing quickly all the
particles hitting its surface, $\eta$ can be very small. For
example, the efficiency parameter of the super-massive BH candidate 
in the Galaxy is estimated to be $\sim 5 \cdot 10^{-6}$~\cite{nara}. 
On the contrary, if all the gravitational energy is released 
as the gas sinks in the potential well of the compact object, 
$\eta = 1 - E_{\rm ISCO}$, where $E_{\rm ISCO}$ is the specific 
energy of the gas particles at the innermost stable circular 
orbit (ISCO) and depends on the metric of the space-time. For 
a Schwarzschild BH, $\eta \approx 0.057$, while for a rotating 
BH $\eta$ can be much higher, up to about 0.42.

If the distance from the compact object is known, $L_{acc}$
can be easily measured. However, an accurate estimate of 
$\dot{M}$ is typically much more problematic and model dependent.
It is instead possible to determine the mean efficiency
parameter of active galactic nuclei (AGN)~\cite{pap0}. From the 
observed hard diffuse X-ray background and a quasar spectral 
energy distribution, one can estimate $u_\gamma$, the total 
contribution of quasar luminosity to the mean energy density 
of the Universe. From the study of the super-massive BH
candidates in nearby galaxies, one can estimate $\rho_{BH}$, 
the mean mass density of BHs in the contemporary Universe.
Under the conservative assumption that these objects acquire 
most of their mass through the accretion process, one divides 
$u_\gamma$ by $\rho_{BH}$, to obtain an estimate of the 
average accretion efficiency $\eta$. Current studies find
$\eta > 0.15$~\cite{pap1,pap3}. There 
are several uncertainties in this value, but 0.15 seems to 
be a reliable lower bound, especially for the most massive 
systems, because it is obtained from a set of conservative assumptions.
An average efficiency around $0.30-0.35$ seems to be a
reasonable estimate~\cite{pap3}. $\eta > 0.15$ is possible for a
rapidly rotating BH with $a_*>0.89$.

{\it Compact objects with non-Kerr quadrupole moment --}
The Manko-Novikov (MN) metric is a stationary, axisymmetric,
and asymptotically flat exact solution of Einstein's vacuum
equation~\cite{mn}. It is not a BH solution, but
it can be used to describe the space-time around a compact
body with arbitrary mass multipole moments. The solution
has an infinite number of free parameters and the full expression
can be seen in Ref.~\cite{noi}, where a few typos present in the
original paper were corrected. Here I consider a subclass
of the MN metric, with only three free parameters: the
mass $M$, the spin parameter $a_*$, and the anomalous
quadrupole moment $q$. The latter is defined by
\be
Q = Q_{\rm Kerr} - q M^3 \, ,
\ee
where $Q_{\rm Kerr} = - a_*^2 M^3$ is the mass quadrupole 
moment of a BH. For $q=0$, we recover exactly the Kerr
metric, while for $q>0$ ($q<0$) the object is more oblate
(prolate) than a BH. The MN solution is written in prolate
spheroidal coordinates and requires $|a_*|<1$, even if
this is not a fundamental limit as in the BH case. However,
at least for small deviations from the Kerr metric, 
compact objects with $a_*>1$ should be unstable~\cite{eb}.

\begin{figure}
\par
\begin{center}
\includegraphics[type=pdf,ext=.pdf,read=.pdf,width=7.5cm]{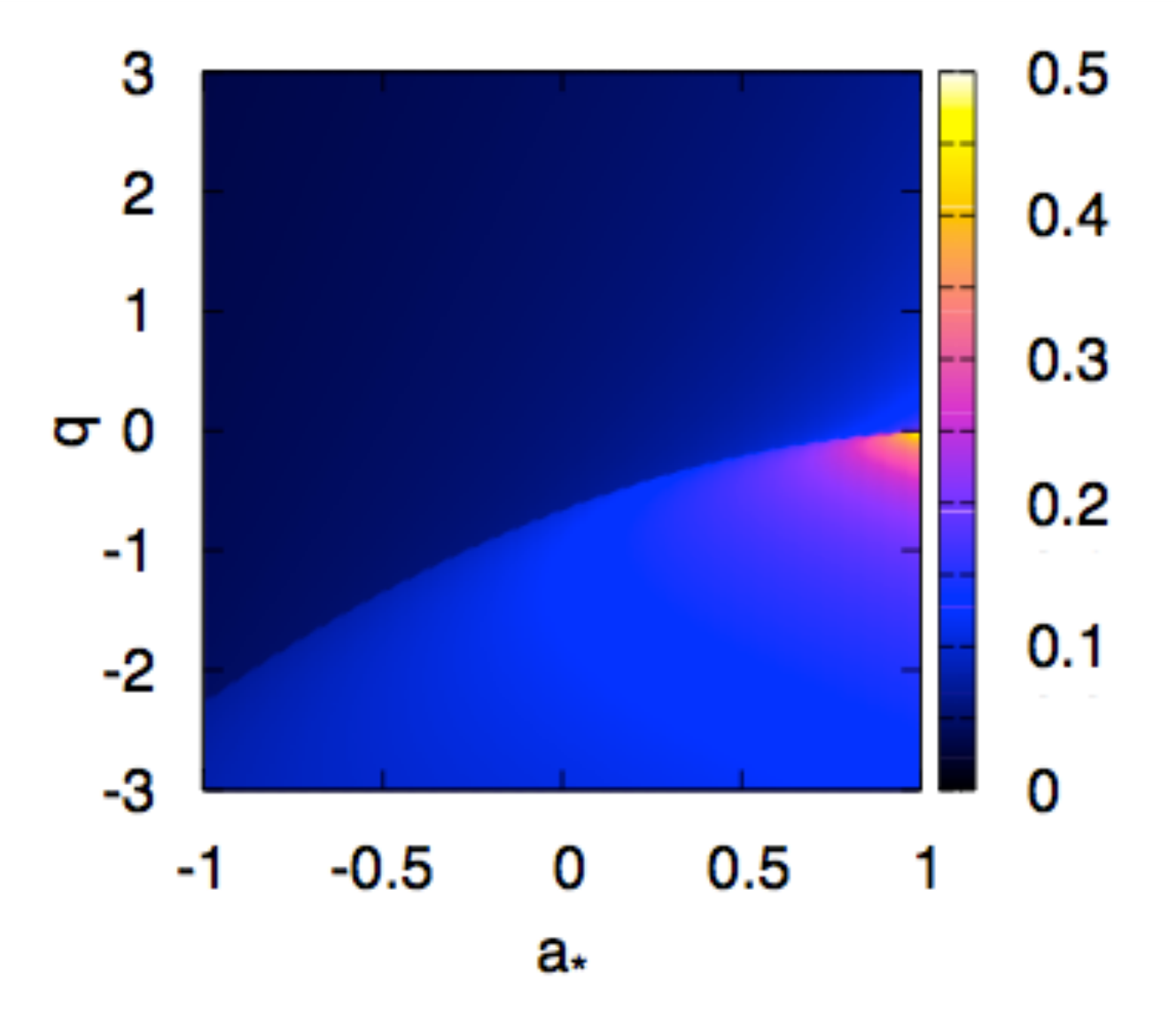}
\end{center}
\par
\vspace{-5mm} 
\caption{Efficiency parameter $\eta$ in the plane $(a_*,q)$.}
\label{f-1}
\end{figure}

The efficiency parameter $\eta$ can be computed as follows.
As in any stationary and axisymmetric space-time, the geodesic 
motion in cylindrical coordinates $(t,r,z,\phi)$ is governed 
by the following equations
\be
\dot{t} &=& \frac{E g_{\phi\phi} + 
L_zg_{t\phi}}{g_{t\phi}^2 - g_{tt}g_{\phi\phi}} \, , \\
\dot{\phi} &=& - \frac{E g_{t\phi} + 
L_zg_{tt}}{g_{t\phi}^2 - g_{tt}g_{\phi\phi}} \, , \\
g_{rr}\dot{r}^2 + g_{zz}\dot{z}^2 &=& V_{\rm eff}(E,L_z,r,z) \, ,
\ee
where $E$ and $L_z$ are respectively the conserved specific
energy and the conserved specific $z$-component of the angular 
moment, while $V_{\rm eff}$ is the effective potential 
\be
V_{\rm eff} = \frac{E^2 g_{\phi\phi} + 2EL_zg_{t\phi}
+ L_z^2g_{tt}}{g_{t\phi}^2 - g_{tt}g_{\phi\phi}} - 1 \, .
\ee
Circular orbits in the equatorial plane are located at the
zeros and the turning points of the effective potential:
$\dot{r}=\dot{z}=0$ implies $V_{\rm eff} = 0$, and
$\ddot{r}=\ddot{z}=0$ requires $\partial_rV_{\rm eff}
= \partial_zV_{\rm eff} =0$. The specific energy turns out
to be
\be
E = - \frac{g_{tt} + g_{t\phi}\Omega}{\sqrt{-g_{tt} 
- 2 g_{t\phi}\Omega - g_{\phi\phi}\Omega^2}} \, ,
\ee
where 
\be
\Omega = \frac{d\phi}{dt} = \frac{-\partial_r g_{t\phi} \pm 
\sqrt{(\partial_r g_{t\phi})^2 - (\partial_r g_{tt})
(\partial_r g_{\phi\phi})}}{\partial_r g_{\phi\phi}}
\ee
is the orbital angular velocity and the sign $+$
($-$) is for corotating (counterrotating) orbits. The orbits are 
stable under small perturbations if $\partial_r^2V_{\rm eff}
\le 0$ and $\partial_z^2V_{\rm eff} \le 0$. 
In this way, one determines the specific energy 
at the inner radius of the disk, $E_{in}$, and the efficiency 
parameter $\eta = 1 - E_{in}$ for a particular choice of the spin 
parameter $a_*$ and of the anomalous quadrupole moment 
$q$\footnote{In the MN space-times, for some $q<0$ one finds
two disconnected regions with stable circular orbits: one 
closer to the object, $r_1<r<r_2$, and another 
for larger radii, $r>r_3$ with $r_3>r_2$. The inner radius of 
the disk is $r_3$, as the orbits in the region $r_1<r<r_2$ 
have larger energy and angular momentum~\cite{noi}.}, see
Fig.~\ref{f-1}.

\begin{figure}
\par
\begin{center}
\includegraphics[type=pdf,ext=.pdf,read=.pdf,width=7.5cm]{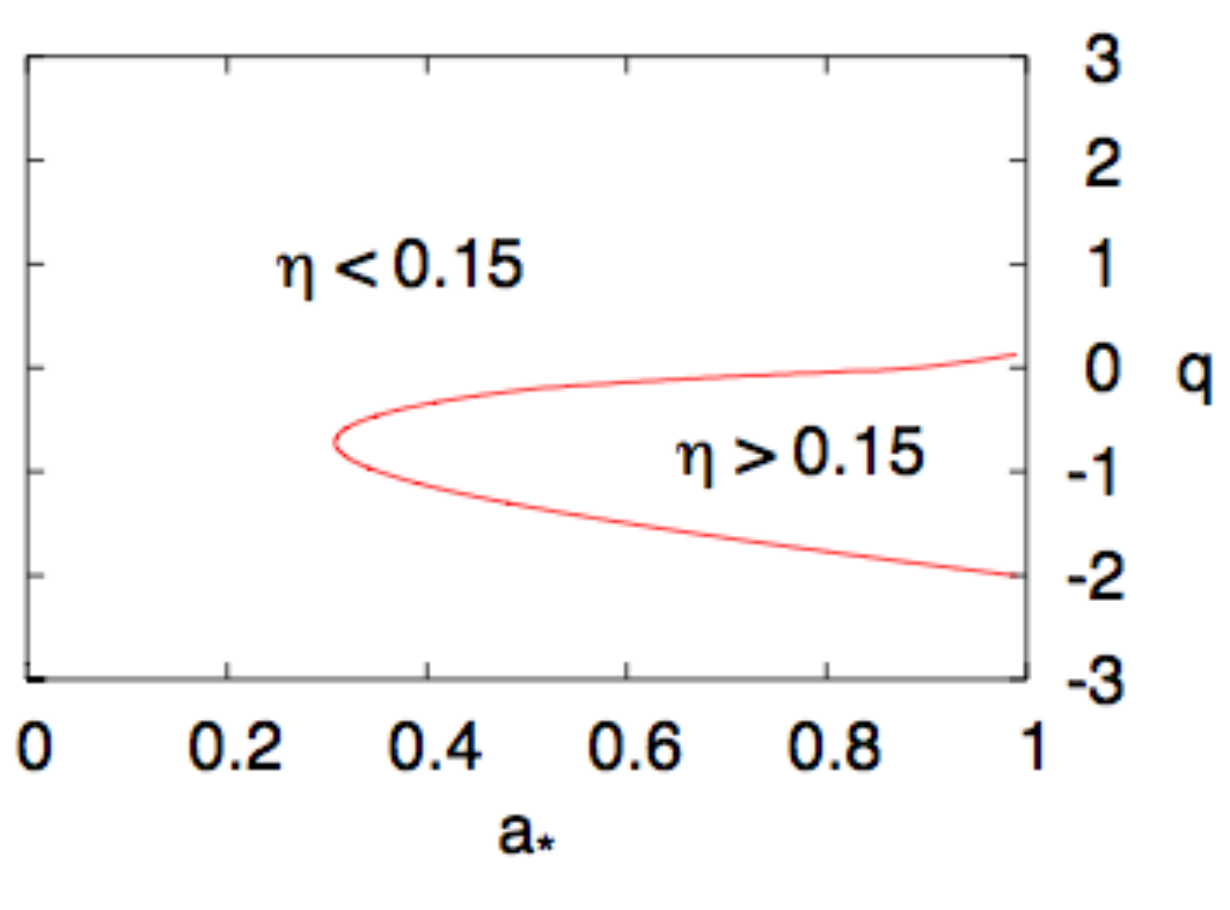}
\end{center}
\par
\vspace{-5mm} 
\caption{Constraint on the mean spin parameter and the mean
anomalous quadrupole moment of AGN from the estimate of their
radiative efficiency. The allowed region is the one with 
$\eta > 0.15$.}
\label{f-2}
\end{figure}

As the mean efficiency parameter of AGN must be larger than 
0.15, one can constrain the mean spin and the mean quadrupole 
moment of these objects. This is done in Fig.~\ref{f-2}, where 
the red curve denotes the boundary between the regions with
$\eta > 0.15$ and $\eta < 0.15$. The constrain on the anomalous 
quadrupole moment $q$ is
\be\label{bound}
-2.01 < q < 0.14 \, .
\ee
If we adopt a more stringent bound on $\eta$, like $\eta>0.20$, 
the constraint on $q$ becomes $-0.96 < q < 0.03$.

{\it Discussion --}
Eq.~(\ref{bound}) provides a constraint on possible deviations
from the Kerr metric around the super-massive BH candidates.
The bound is much weaker for negative values of $q$ because
in these space-times either the inner radius of the disk and
the specific energy at a given radius are usually smaller than 
the cases with $q>0$. It is clear that the super-massive BH 
candidates must be objects very different from a compact
body made of ordinary matter. For instance, the quadrupole 
moment of a neutron star is thought to be well approximated 
by the following expression
\be
Q = - (1 + \tilde{q}) a^2_* M^3 \, ,
\ee
with $\tilde{q} \approx 1 - 10$ independent of $a_*$, 
according to the matter
equation of state and the mass of the body~\cite{laa}.

The constraint in Eq.~(\ref{bound}) relies on the assumption that 
the mass of these objects is conserved during mergers. While 
this is a reasonable approximation for BHs in general relativity, 
we cannot say anything 
in the case of compact objects with unknown internal structure. 
If a substantial fraction of their mass were lost during merger, 
for instance through the emission of gravitational waves, the 
bound would be weaker, as the energy radiated in the accretion 
process would come from a larger amount of accreted mass. To
obtain Eq.~(\ref{bound}), I also assumed that the disk is on the
equatorial plane. As explained in~\cite{ss}, this is justified
by the fact that the timescale of the alignment of the spin of 
the object with the disk is much shorter than the time for the 
mass of these objects to increase significantly.

I would like to warn the reader that the estimate of $\eta$ in 
Fig.~\ref{f-1} and the allowed region in Fig.~\ref{f-2} for 
the spin and the anomalous quadrupole moment inevitably partially 
depend even on 
the higher order moments of the space-time. The latter are
less and less important, but they are not completely negligible. 
This can be easily understood by noticing the difference in the
constraint on the spin parameter $a_*$ between a BH with $q=0$ 
and a generic object with $q \neq 0$. For a BH, an efficiency 
parameter $\eta$ larger than 0.15 requires $a_* > 0.89$. For 
$q \neq 0$, this bound relaxes to $a_* > 0.30$, see Fig.~\ref{f-2}.
This problem is present in any estimate of a quadrupole moment 
and therefore the future comparison of two limits on $q$ obtained 
from different arguments or with different metrics deserves some
attention.

{\it Conclusions --}
There are not yet direct observational evidences that the
super-massive objects at the center of many galaxies are the
BHs predicted by general relativity, while recent theoretical
arguments suggest that the final product of the gravitational
collapse of matter may be quite different from what it is 
usually thought~\cite{new}. The BH hypothesis can be tested 
by measuring at least three non-trivial moments of the 
gravitational field of these objects, as in the case of a BH 
all the moments depend on the mass $M$ and the spin $J$ in 
a very specific way. There are several works in the literature
discussing how this is possible with future experiments,
but so far there are no constraints on the nature of these
objects. For example, the future gravitational wave detector 
LISA will be able to measure the quadrupole moment of the
super-massive BH candidates with a precision at the level of 
$10^{-2} - 10^{-4}$ (see the third paper in~\cite{gw}). 
In this letter, I considered the current estimate of the mean 
radiative efficiency of AGN and I was able to constrain the 
anomalous quadrupole moment $q$ of these objects. The 
bound I obtained is $-2.01 < q < 0.14$.

Lastly, let us notice that the maximum radiative efficiency 
for a BH is $\eta \approx 0.42$ when $a_* = 1$. A very 
fast-rotating object with $q$ a little bit smaller than 0 can 
have a higher efficiency parameter. This implies, at least 
in principle, that the argument used in this paper may also 
rule out the Kerr BH hypothesis in the case of the discovery 
of an object with an efficiency parameter larger than the 
one that can be expected for a BH.

\begin{acknowledgments}
I would like to thank Sergei Blinnikov, Stefano Liberati, 
Charles Steinhardt, and Naoki Yoshida for useful discussions
and comments. This work was supported by World 
Premier International Research Center Initiative (WPI Initiative), 
MEXT, Japan, and by the JSPS Grant-in-Aid for Young Scientists 
(B) No. 22740147.
\end{acknowledgments}


\end{document}